# Corridor for new mobility Aachen - Düsseldorf: Methods and concepts of the research project ACCorD

**Laurent Kloeker[1*], Amarin Kloeker[1], Fabian Thomsen[1], Armin Erraji[1], Lutz Eckstein[1], Serge Lamberty[2], Adrian Fazekas[2], Eszter Kalló[2], Markus Oeser[2], Charlotte Fléchon[3], Jochen Lohmiller[3], Pascal Pfeiffer[4], Martin Sommer[5], Helen Winter[6]**

1. Institute for Automotive Engineering, RWTH Aachen University, Germany
2. Chair and Institute of Highway Engineering, RWTH Aachen University, Germany
3. PTV Planung Transport Verkehr AG, Germany
4. e.GO MOOVE GmbH, Germany
5. Ford-Werke GmbH, Germany
6. City of Aachen, Germany
*laurent.kloeker@ika.rwth-aachen.de

**Abstract**

With the Corridor for New Mobility Aachen - Düsseldorf, an integrated development environment is created, incorporating existing test capabilities, to systematically test and validate automated vehicles in interaction with connected Intelligent Transport Systems Stations (ITS-Ss). This is achieved through a time- and cost-efficient toolchain and methodology, in which simulation, closed test sites as well as test fields in public transport are linked in the best possible way. By implementing a digital twin, the recorded traffic events can be visualized in real-time and driving functions can be tested in the simulation based on real data. In order to represent diverse traffic scenarios, the corridor contains a highway section, a rural area, and urban areas. First, this paper outlines the project goals before describing the individual project contents in more detail. These include the concepts of traffic detection, driving function development, digital twin development, and public involvement.

## Introduction

The research and development of automated and connected driving are progressing at high speed. Cost-intensive efforts are being made to develop new and further methods to increase the scope of automated driving functions, for example, driver-assisting features and safety-relevant functions. However, a broad database of traffic situations is often lacking for the development and, in particular, the validation of automated driving functions. Therefore, traffic data collection is of particular importance to achieve these goals.

By creating an integrated development environment, this data can be easily and cost-efficiently used for further development. The basis for this environment is formed by multimodal sensors, which record raw data from real traffic and convert the raw data into anonymized traffic objects. The sensors will be deployed in three designated test fields to provide the capability of recording the entire traffic for each field. These test fields cover a wide range of domains in real traffic, namely urban traffic, rural roads as well as highways. Besides, testing of vehicle-to-everything (V2X) communication is taking place on parts of the test fields. The traffic data collection is complemented by the development of a digital twin, which further increases the number of traffic situations by means of behavioural models of road users parameterized based on the collected traffic data.

## Related work

Digital test fields for automated and connected driving in real traffic are represented in almost all of Germany's federal states. However, the focus of these test fields mostly relates to V2X communication and only rarely includes additional traffic recording by infrastructure sensors. Nevertheless, the recording of traffic events and the collection of data as a basis for the development and safeguarding of current and future sensor technologies and automated driving functions as well as the provision of test infrastructures are currently being applied and implemented in some research projects.

Corridor for new mobility Aachen - Düsseldorf: Methods and concepts of the research project ACCorD

For inner-city traffic detection, the Application Platform Intelligent Mobility (AIM) in Braunschweig [1], consisting of a research intersection for real-time detection of road users, prediction and classification, and trajectory and scene video detection, are particularly relevant in this context. In the area of providing an extra-urban test field and traffic detection in real traffic, the activities Test Field Lower Saxony [2], and Providentia++ in Bavaria [3] can be highlighted in particular in this context.

The Test Field Lower Saxony includes urban test areas as well as highways and rural roads. Mobile and fixed masts equipped with high-resolution stereo cameras and LiDAR sensors are used for traffic detection. The main focus of traffic detection is on a section of the A39 highway.

The test field of the Providentia++ research project is based on the predecessor project Providentia and focuses on the digital test field highway on the A9. Camera and radar sensors on gantries are used to identify and anonymize objects. By fusing the sensor data, a digital twin is created in real-time that broadcasts real-time information to vehicles, supports behavioural decisions of connected and automated vehicles, and visualizes traffic in a simulation environment.

**Method**

*Project objective*

The research project ACCorD aims to create an integrated development environment for systematically testing and validating connected and automated vehicles (CAVs). Multiple important parts incorporating public test fields as well as simulations are linked together to achieve this goal. A general overview of the project scope is provided in Figure 1.

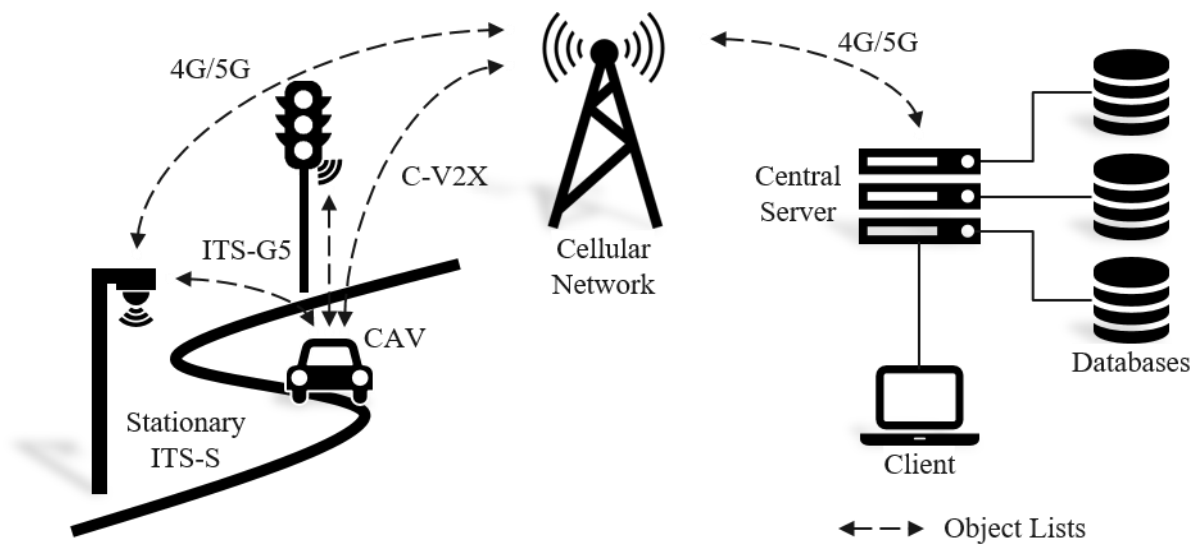

**Figure 1 - Overview of the ACCorD project scope.**

First, three test fields consisting of stationary ITS-Ss are developed and constructed to permanently record all road users with high accuracy and precision. The test fields are located in an urban and rural environment as well as on a highway to cover a large variety of relevant traffic scenarios. An IT infrastructure is implemented for the communication between the ITS-Ss and a central server, which extracts trajectories of all perceived objects. On the one hand, the recorded data is broadcasted live to test vehicles in the respective areas to both, supplement and validate their own environment sensor systems. This allows investigating prototypic automated driving functions and their dependency on precise environment perception. On the other hand, the offline comparison between trajectories registered by the ITS-Ss and CAVs enables the validation and optimization of the latter's systems. On an additional site, traffic lights are also equipped with V2X components to test their interaction with automated vehicles. All acquired data is stored in a central database for further research and development

Corridor for new mobility Aachen - Düsseldorf: Methods and concepts of the research project ACCorD

for which naturalistic traffic data has shown to be a valuable basis. Scenarios of particular interest are extracted and investigated in detail as they form a key component for the comprehensive validation of automated driving functions.

Second, a digital twin of the test fields is created to visualize live trajectories and to test driving functions in simulations based on real traffic data. For this objective, the general human driving behaviour is researched to derive a behavioural model for improving the simulation of lateral vehicle movements and lane changes in the field of microscopic traffic flow simulation. These models are implemented in the PTV Vissim simulation software. The improvement of microscopic traffic flow simulation regarding vehicle lateral control is necessary for connection with the increasing automation and connectivity of motor vehicle traffic to better represent driving characteristics and quantify optimization potential. The use of the vehicle and infrastructure data recorded within the project for the development and parameterization of behaviour models of road users is an essential contribution to the development of the digital twins.

*Traffic detection concept*

Each corridor will be equipped with stationary ITS-Ss at a distance of 60 to 100 m to each other. The equipment of each ITS-S includes two 128-layer LiDAR sensors, two 4K cameras, a V2X Road Side Unit (RSU), a 4G communication module, and a computation unit to process the data. The two LiDARs are oriented to cover the immediate area within 50 m of each ITS-S. This ensures that the entire corridor is covered by LiDARs. The cameras, on the other hand, are oriented so that one camera from each ITS-S covers the section in front of an adjacent ITS-S. The goal of this configuration is to ensure that, for one ITS-S, a road user obscured in the near section can still be detected by the camera of a neighbouring ITS-S.

The sensors generate a raw data volume of about 100 MB/s per station. Since this amount of data is too large to be streamed unprocessed to a central server, the raw data is processed directly at each ITS-S into an object list. For this purpose, each ITS-S has a computing unit that is powerful enough to extract all road users from the raw data with high accuracy in real-time using modern computer vision and machine learning algorithms.

The object lists generated in this way are then streamed to a central server via the 4G mobile network. There, the object lists of all ITS-Ss are merged into a digital image of the corridors. Overall, each road user is to be detected with an accuracy of less than 10 cm.

There are generally two use cases for the generated data. First, the transmission of the data to CAVs via V2X. For this purpose, two channels are available: C-V2X and ITS-G5. While for C-V2X the data is streamed directly from the central server, for ITS-G5 the data is fed back to each ITS-S and distributed from there via the RSU. Second, the data is stored on a database for multiple uses at a later time. Possible applications are the creation of a traffic dataset, the extraction of interesting scenarios, or the subsequent validation of tested driving functions.

*Infrared camera-based ITS-S*

In addition to the above described stationary ITS-Ss, six stationary infrared camera-based ITS-Ss will be installed on the highway test field at a distance of appr. 100 m to each other to collect traffic data during the night and in poor weather conditions. Each unit is composed of a thermal imaging traffic surveillance camera along with a computing unit to extract traffic data. The obtained trajectories will be stored in the above described central server for later analysis.

Infrared cameras detect the temperature difference between the street and vehicles, which makes this ITS-S feasible for continuous traffic data collection both during the night and in poor weather conditions. At the same time, infrared cameras ensure data privacy. Since the licence plate and the characters on it have the same temperature as the vehicle, the licence plate remains invisible in case of recording [4]. Furthermore, in many cases there is no electricity available on the side of motorways, thus infrared camera-based ITS-S can provide an easy, affordable and sustainable solution through their low power of max. 100 Watt to build test sites.

The described infrared camera-based ITS-S was previously tested within the European research project MeBeSafe. The project aimed to influence drivers directly through a complex nudging system built in the infrastructure, based on real-time trajectory data collection via ITS-S using computer vision algorithms developed at ISAC [5].

In ACCorD, we analyse the main challenges in obtaining valuable microscopic traffic data for CAVs via infrared camera-based ITS-S. We study the usability and limitation of the system to determine the reachable level of data quality regarding resolution, classification, accuracy, precision and latency. The result of the work is an analysis, in what terms thermal imaging cameras can meet the minimum requirements to provide useful data for CAVs, regardless of the weather conditions. In addition, the collected traffic data can be used for further analysis e.g. for microscopic traffic safety analysis [6].

*Microscopic traffic flow simulation and the improvement of lateral behaviour models*

The necessary use of vehicle, infrastructure, and wireless technology for highly and fully automated driving is very cost-intensive. Furthermore, the technology for automation and communication functions in the vehicle as well as at the roadside is currently still under development. Therefore, simulations can support the development, identifying possible weak points and their causes as well as validating measures to positively influence traffic flow. Changes in driving functions and the increasingly complex vehicle interactions through automation and connectivity can also be depicted and analysed using traffic flow simulation.

For the simulation of connected and automated vehicles within the microscopic traffic flow simulations, the individual driving behaviour of conventional vehicles must be represented as realistically as possible. Over the last decades, corresponding behaviour models for the longitudinal guidance of conventional vehicles have been developed [7]. Thus, acceleration, braking, and distance behaviour are modelled in detail. Initial approaches also exist for modelling the overtaking behaviour [8], [9]. However, the models for lateral driving behaviour and lane change behaviour were neglected in this context. The background for this is that the existing traffic infrastructure is implicitly sufficiently dimensioned, based on the experience of the past decades in connection with motor vehicle traffic, to enable these driving processes for conventional vehicles. Therefore, there was no need to reproduce this driving behaviour more accurately.

Behaviour models for the longitudinal guidance of connected and automated vehicles have already been adapted [10]. However, with the increasing market entry of vehicles with automated driving functions, the question arises as to whether and which requirements exist for these driving functions due to lateral driving behaviour and lane change behaviour and which technical solutions are best suited to meet these requirements. For example, connected automated driving functions, which lead to the assumption of more efficient lane changes, can increase the capacity of the traffic infrastructure. These capacity increases could be quantified by further development of microscopic traffic flow simulation and ultimately lead to a more efficient dimensioning of the road infrastructure.

The few studies that exist reveal the existence of different phases in lane changes, starting with the decision for/against a lane change and ending with the execution of the lane change. Each phase corresponds to a different behaviour model [11], [12]. Compared to the behavioural models for longitudinal movement, there are very few models for lateral movement and lane changes, because these behavioural models require significantly higher accuracy of empirical data. The collection of such data is very costly and human behaviour when changing lanes differ regionally due to different traffic rules and cultural conditions.

To derive general correlations for the description of lane change behaviour and the behaviour regarding further lateral vehicle movements, vehicle data collected by the project partners within the framework of the project will be analysed and bundled into so-called trajectories of individual vehicles. These data

Corridor for new mobility Aachen - Düsseldorf: Methods and concepts of the research project ACCorD

will be evaluated concerning possible correlations which can be used for a later formulation of a behavioural model approach.

In parallel, digital twins are being set up with PTV Vissim to serve both the calibration of the established behaviour models and the further development of the automated and connected driving functions to be carried out.

*City application scenario*

In the city application scenario, the e.GO Mover is used as an automated and electrically operated shuttle (people mover). As a mobility solution, the people mover is intended to transport people along a predefined route without the need for a driver and allows them to enter and exit the vehicle at defined points. In a further development stage, this technology is to be used more flexibly, i.e. independent of routes and stops (on-demand), to increase the efficiency of the solution and to optimize the mobility options. In both cases, the operating space for which the system is developed is known (ODD: Operational Design Domains) and geographically limited, so that the challenge of perceiving the environment and the action of the people mover in the traffic context can be supported by the use of application and location specific maps and reference data (planning speed, location specific behaviour, etc.).

For ACCorD, the e.GO Mover is aimed to perform the manoeuvre "operating a bus stop" since this is especially relevant for a people mover. Especially due to the focus of the research landscape on automated private vehicles, little attention has been paid to this application case so far. For this reason, the manoeuvre itself represents a special case that must be researched and systematically integrated into the validation methodology.

Typical use cases in research for automated driving include highway driving or valet parking. Even more than in those applications, for the bus stop manoeuvre, strong requirements are given to the localization. Since people get close to the vehicle, especially in the vicinity of the bus stop, and since the vehicle moves towards the structural edge of the road (curb, etc.), this manoeuvre places special demands on the relative localization to the corresponding objects. Importantly, the distance between the people mover and the bus stop must be small so boarding is as safe as possible ("mind the gap"). The people mover will implement this manoeuvre in a map- and perception-based manner so that it can be used as a research object for the entire project.

The automated driving system (ADS) consists of a perception system, a planning system, and an acting system. Defining subsystems with precise system boundaries and interfaces paves the road for robust validation methods. The perception system detects the relevant aspects of the vehicle's environment and interprets the current driving situation. In the bus stop manoeuvre, perceived objects (landmarks) will be used for localization, while road participants, such as pedestrians, cyclists, and animals will influence the planned trajectory. In addition to the perceived environment information, the people mover has access to a digital representation of the static environment (digital map) which contains the drivable corridor (predefined navigation), the speed limit profile, and other information. The planning systems will derive a safe and comfortable driving strategy and will plan a collision free and comfortable trajectory for the next seconds (guidance task). The speed of the trajectory will be adapted predictively according to the manoeuvres (reacting on objects, stopping at a bus stop, etc.).

Additionally, an overall validation methodology, which, supported by the measuring stations, is intended to provide a validation basis for the safety of the overall system. Here it is important that an independent and complete database is generated (reference data). By this means, within the acting system, the controller compares the actual sensed position with the desired trajectory position and the actual velocity with the desired trajectory velocity and eliminates the deviation by requesting the brake, the electric drive and the electric steering (stabilization task) to move the vehicle along the planned trajectory. The path and the velocity profile are followed accurately, while disturbances such as side wind and road slope are compensated. The controller is implemented by a state of the art feed-forward and closed-loop approach. These types of data (position, orientation, velocity) will be used to generate the database (validation data), which will be then validated supported by the data provided from the measuring stations (reference data). Camera sensors pointing forwards perceive static and dynamic objects and classify these objects. Also, the camera processing estimates the relative position of the object, which has high accuracy in vertical (sensor y-direction) and horizontal (sensor z-direction) direction but low accuracy in the distance (sensor x-direction). LiDAR sensors with one or more layers create point clouds with highly accurate distance and angle information, which are transformed in the vehicle coordinate system. However, the density of the point clouds is lower than camera resolution and instead of colour information, a reflectivity value is given by the sensor. For validation purposes, requirements for the measuring stations that will support the overall project are derived. In the first step, the data will be validated offline.

In the initial phase, the bus stop manoeuvre will be performed without influence from other road participants. This also has the advantage that the actions the safety driver has to conduct when an intervention is needed, are limited and well known by the driver. For evaluating the manoeuvre in a more realistic scenario the effect of external participants is to be investigated in future. Critical scenarios with, among others, crossing pedestrians and cyclists behind the vehicle should be taken into consideration. Ultimately, a concept for remote operation of the automated driving people mover will be developed and evaluated. In this case, the safety driver is not physically present in the vehicle but will be able to intervene via remote control. In remote operation, it must be ensured that the system can handle all situations in a safe manner. With the implementation of the remote control, the aspect of cybersecurity must be analysed and assessed. For the safety of the passengers and other road participants, an unauthorized intervention must be prevented and the connection to the safety driver must be robust. In addition, the behaviour of the people mover, in case of losing the remote connection, must be defined.

*On-road validation of a V2I based traffic signal co-pilot in the connected traffic light corridors*

In the last decades, vehicle manufacturers have continuously introduced new driver assist features on their way to higher automation and especially in the last years vehicles became more and more connected. While in 2018 only 8% of all cars worldwide were connected, in 2023 approximately 24% will be connected [13]. Thus, using information coming from connected infrastructure and vehicles (V2X or Car2X communication) will be an important part of future driver assistance systems and automated vehicles respectively. The domains connectivity and automation, which in the past were predominantly separated, will come together in the future. This supports the driver to reduce stress, especially as the vehicle, in contrast to the driver, knows for example about future traffic light state transition timing, this way the vehicle can act more proactively.

Within the project, Ford develops a vehicle-to-infrastructure (V2I) based traffic signal co-pilot which can adapt the vehicle's speed according to the traffic light's status. Ideally, the vehicle speed is controlled to smoothly pass the traffic light at green. However, in case this is neither reasonable nor possible (e.g. too low speeds required), the vehicle stops in front of a red traffic light and will resume after the transition from red to green. Compared to the state of the art, there are several challenges to be addressed. The timing forecast of traffic dependent or actuated traffic lights can be inaccurate and characterized by sudden and significant changes, which requires a control strategy being capable of accommodating such forecast uncertainties. A way to handle low confidence timing information needs to be developed to ensure a safe and smooth vehicle control. Inaccuracies of input information (e.g. vehicle position, traffic light timing) will challenge the trade-off between a conservative behaviour on the one hand and a more aggressive natural driver-like behaviour on the other hand. That said, we have identified the following research questions which shall be answered during the project:

1. Is the reliability of V2I information sufficient to be used for automated longitudinal control?
2. Are both direct communication and network-based communication delivering the required performance, e.g. in terms of range, connection stability & latency?
3. How does the functionality impact the overall traffic-flow?
4. Can the functionality work well under all conditions or only in a subset (e.g. only rural conditions)?

Corridor for new mobility Aachen - Düsseldorf: Methods and concepts of the research project ACCorD

In the experimental setting, the communication will be based on direct communication as well as network-based (4G/5G) communication. A test vehicle providing ACC stop&go functionality is used and modified for validation and testing purposes. A stepwise approach is pursued to test and validate the feature: The early development stage focuses on software-in-the-loop (SiL) tests using artificial scenarios, while in the next step the feature will be evaluated using a test vehicle being operated in the Aldenhoven Testing Center (ATC). Once the required robustness is achieved, final testing and validation will be performed within the connected traffic light corridors on public roads. Of course, the impact of such a feature on the overall traffic flow is important to its acceptance. For this reason, a co-simulation of the feature simulation model together with the macroscopic traffic flow simulation software PTV Vissim will be considered as well.

One of the two connected traffic light corridors being built up in the project is located in a city environment equipped with primarily non-adaptive timing but with potential trigger events like an emergency vehicle priority request. The speed limit is 50 kph. The other connected traffic light corridor is set up in a rural environment, dominated by adaptive traffic light timing (traffic dependent) and speed limits of mostly 70 kph and higher. This setup provides the opportunity to gain knowledge about the applicability under significantly different operating conditions. In essence, a corridor for connected and automated driving is crucial to holistically answer all research questions.

*Living Labs: Collaboration for creation and urban innovation through active citizens' participation*

Over the last couple of years, the Aachen region has become one of the most innovative regions for Mobility 4.0 and the development of new and innovative technology regarding connected and automated mobility.
For a sustainable regional development, the implementation of new and future-driven technologies is key to not only face and overcome the challenges of climate change but also to provide further economic development and sustain a liveable city for everyone.

As the city administration, the focus is not only on supporting the local economy by enabling the implementation of innovative technologies but also to actively involve the citizens in the digital urban transformation. ACCorD aims to create a living lab within the city that accompanies both of these components.

Living labs are often used in innovation processes to ensure the active participation of citizens as well as enterprises, research and politics. Those labs create a user-centred and open innovation space that integrates ongoing research and innovation technologies into a city. The collaboration between civil society and researchers creates a highly creative and innovative environment for ideas to the challenges of the future [14].

In order to create acceptance for automated and connected driving among the citizens of Aachen the project establishes a 'Mobility Store' as a presentation space where information will be provided through objects, media and information material on the technical planning of the project.

The Mobility Store aims to establish a place of open and cooperative communication and collaboration between all stakeholders in the city. The Mobility Store is part of the living lab and co-creation space 'OecherLab' which is located in the city centre of Aachen. The living lab 'OecherLab' is the foundation of all citizens' participation processes. It provides a variety of digital and innovative tools and methods for the active involvement of citizens as well as government, companies and research institutes in innovation processes in Aachen. With the help of a digital game and query methods, user acceptance and feedback are created leading to a better understanding of the implemented technologies.
Since the innovative technologies are being tested and implemented in public spaces, an assessment and support of the safety-related and legal components are crucial for the validation. The city of Aachen provides support with legal, planning and implementation processes but also actively involves the citizens in the transformation of the infrastructure creating a living lab and shaping the future of connected and autonomous driving in Aachen.

Corridor for new mobility Aachen - Düsseldorf: Methods and concepts of the research project ACCorD

**Conclusion and outlook**

In our work, we have shown that the establishment of an integrated test environment in real traffic is an essential component for future research and development in the field of connected and automated driving. By covering all domains (urban, rural, highway) with stationary ITS-Ss, testbeds are created that provide valuable results beyond the project timeframe. Only through a cooperative interaction of data providers (ITS-Ss) and data receivers (CAVs, digital twin) challenges in the development of automated and connected vehicles can be identified and solved.

As the conceptual design phase is completed at this stage, the next step consists of setting up the test fields and deploying the research vehicles. Meanwhile, software developments from all partners will be continuously continued and optimized with the help of the collected data from the ITS-Ss and vehicles. The involvement of citizens during the entire project period creates a bidirectional exchange that generates not only information but also ideas and suggestions on both sides. In the end, the collected data will be finally analysed and evaluated. Valuable conclusions will also be drawn for future digital test fields, the number of which is likely to increase in the future.

**Acknowledgement**

The research leading to these results is funded by the Federal Ministry of Transport and Digital Infrastructure (BMVI) within the project "ACCorD: Corridor for New Mobility Aachen - Düsseldorf" (FKZ 01MM19001). The authors would like to thank the consortium for successful cooperation.

**References**

1. Schnieder, L., & Lemmer, K. (2012). Anwendungsplattform Intelligente Mobilität-eine Plattform für die verkehrswissenschaftliche Forschung und die Entwicklung intelligenter Mobilitätsdienste. Internationales Verkehrswesen, 64(4), 62-63.

2. Köster, F., Mazzega, J., & Knake-Langhorst, S. (2018). Automatisierte und vernetzte Systeme Effizient erprobt und evaluiert. *ATZextra*, *23*(5), 26-29.

3. Krämmer, A., Schöller, C., Gulati, D., & Knoll, A. (2019). Providentia-a large scale sensing system for the assistance of autonomous vehicles. *arXiv preprint arXiv:1906.06789*.

4. Berghaus, M. et al. (2019). Infrastructure measures (Deliverable 3.3).

5. Köhler, A.-L. et al. 2019. Report Infrastruktur Measures (Deliverable 3.2) Retrieved from https://www.mebesafe.eu/deliverables/

6. Fazekas, A., Hennecke, F., Kalló, E., Oeser, M. (2017). A Novel Surrogate Safety Indicator Based on Constant Initial Acceleration and Reaction Time Assumption. *Journal of Advanced Transportation*, vol. 2017, Article ID 8376572, 9 pages.

7. Wiedemann, R. (1974). Simulation des Straßenverkehrsflusses. Technical report 8, *Institut für Verkehrswesen, University of Karlsruhe, Germany*.

8. Mocsári, T. (2009). Analysis of the overtaking behaviour of motor vehicle drivers. *Acta Technica Jaurinensis*, *2*(1), 97-106.

9. Asaithambi, G., & Shravani, G. (2017). Overtaking behaviour of vehicles on undivided roads in non-lane based mixed traffic conditions. *Journal of traffic and transportation engineering (English edition)*, *4*(3), 252-261.

10. Zeidler, V. (2018), Evaluierung und Weiterentwickung der Simulation autonomen Fahrverhaltens mit der Software PTV Vissim, *Master Thesis, Institut für Verkehrswesen, Karlsruhe Institute of Technology, Germany.*

11. Sparmann, U., Hrsg. (1978). Spurwechselvorgänge auf zweispurigen BAB–Richtungsfahrbahnen. *Forschung Straßenbau und Straßenverkehrstechnik*, Heft 263, Bundesministerium für Verkehr,

Corridor for new mobility Aachen - Düsseldorf: Methods and concepts of the research project ACCorD

Abt. Straßenbau, Bonn.

12. Kesting, A.; Treiber, M., Helbig, D. (2006). MOBIL: General Lane-Changing Model for Car-Following Models, Dresden. Retrieved from http://www.mtreiber.de/publications/MOBIL_TRB.pdf.

13. Caeser, M., Mehl, R., Tschoedrich, S. et al. (2019), Connected Vehicle Trend Radar, *Capgemini invent*, Issue 1 2019.

14. Almirall, E., & Wareham, J. (2011). Living Labs: arbiters of mid-and ground-level innovation. *Technology Analysis & Strategic Management*, *23*(1), 87-102.